\documentstyle[eqsecnum,floats,aps,prb,epsfig,twocolumn]{revtex}

\begin{document}

\wideabs{
 
\title{ 
Intra- site $4f-5d$ electronic correlations  
in the quadrupolar model of the 
$\gamma$-$\alpha$ phase transition in Ce. }

\author{A.V. Nikolaev$^{1,2}$ and K.H. Michel$^1$ }

\address{
$^1$Department of Physics, University of Antwerp, UIA, 
2610 Antwerpen, Belgium \\
$^2$Institute of Physical Chemistry of RAS, 
Leninskii prospect 31, 117915, Moscow, Russia 
}

\maketitle 

\begin{abstract} 
As a possible mechanism of the $\gamma-\alpha$ phase transition
in pristine cerium a change of the electronic density from
a disordered state with symmetry $Fm{\bar 3}m$ to an ordered
state $Pa{\bar 3}$ has been proposed.
Here we include on-site and inter- site electron correlations
involving one localized $4f$-electron and one conduction $5d$-electron
per atom.
The model is used to calculate the crystal field of 
$\gamma$-Ce and the temperature evolution of the
mean-field of $\alpha$-Ce. 
The formalism can be applied to crystals where quadrupolar
ordering involves several electrons on the same site.
\end{abstract}

%
%


}

\section {Introduction} 
\label{sec:int}

Elemental solid cerium is known to undergo structural
phase transitions. \cite{Kos,McMah} In the
pressure-temperature phase diagram of Ce the puzzling 
long-standing problem is to understand the
apparently isostructural transition between the cubic
$\gamma$ and $\alpha$ phases. \cite{Kos,Mal}
An isostructural phase transformation can not be ascribed to a
condensation of an order parameter and therefore can not 
be explained by the Landau theory of phase transitions.
In the past, several models and theories have been suggested to
address this problem. \cite{Joh,Joh1,Sva,Jar,Laeg,All2,Lav,Gun,Bic}
Among them, a Mott like transition for 4$f$ electrons \cite{Joh,Joh1}
and a Kondo effect based approach \cite{All2,Lav,Gun,Bic} 
are the competing ones.
Also, new computational schemes have been applied to the problem using
dynamical mean-field theory combined with the local
density approximation. \cite{Zol}

Under pressure above 5 GPa $\alpha$-Ce becomes unstable
and transforms first to a crystal with $C2/m$ or $\alpha$-U
space symmetry \cite{McMah} ($\alpha'$-Ce) and then to a body-centered
tetragonal (bct) structure ($\alpha''$-Ce) above 12 GPa. \cite{Endo} 
This series of transformations can not be explained by invoking
the concepts of the 4$f$ localization-delocalization transition 
or Kondo volume collapse models and indicates that there are
anisotropic interactions present in the $\alpha$ phase of cerium. 
Such electron interactions can be of quadrupolar origin
that are known to drive symmetry lowering phase transitions
in many lanthanide and actinide compounds. \cite{Mor1}

Recently, the isostructural character of the $\gamma-\alpha$ 
phase transition has been
questioned by Eliashberg and Capellmann. \cite{EC} 
They suggest that $\alpha$-Ce should have a distorted fcc structure.
Independently, the present authors have put forward a theory
of quadrupolar ordering in cerium. \cite{NM1,NM2}
There, it was proposed that the $\gamma-\alpha$ transformation is not
really isostructural. Rather, it was associated with hidden
electronic degrees of freedom. \cite{Eb}
In our previous work, Refs.\ \onlinecite{NM1,NM2}, 
we have suggested that the symmetry change
is from $Fm{\bar 3}m$ to $Pa{\bar 3}$. 
This symmetry lowering is a special one. Although accompanied
by a lattice contraction, it conserves
the fcc structure of the atomic center of mass points
(cerium nuclei) and is solely due to orientational
order of electronic densities. 
Such a scenario reconciles
the $\gamma-\alpha$ transformation with the Landau theory
of phase transitions. Our considerations for cerium have been
inspired by the theory \cite{Mic1} of orientational phase
transition in solid C$_{60}$ where a similar space symmetry change
($Fm{\bar 3}m \rightarrow Pa{\bar 3}$) occurs at 255 K at room
pressure. \cite{C60}

The present article is a continuation of our approach to
the problem of the $\gamma-\alpha$ transition in Ce
based on the technique of multipolar
interactions between electronic densities of conduction and
localized electrons in a crystal. \cite{NM1,NM2}
Our second motivation is to extend our initial method, Ref.\ \onlinecite{NM1}, 
for the case when two electrons ($f-$ and $d-$) are at the same site
of cerium and all intra- site interactions 
(including the on-site exchange) between them are 
taken into account. Our treatment of intra- site correlations
is closely related with the method used by Condon and Shortley
for many electron states of atoms. \cite{CS} 
Provided that the average number of electrons per site
is conserved this method is exact for intra- site correlations and goes beyond
the usual self-consistent-field approach \cite{LAPW,LMTO,LMTO1} 
employed by band structure calculations.

Besides the problem of the $\gamma-\alpha$ phase transition in Ce,
the microscopic method can be applied further \cite{rem3} to describe 
quadrupolar ordering and to perform crystal field
calculations of many $f$ electrons on the same site.
There are numerous compounds exhibiting quadrupolar ordering
at low temperatures \cite{Mor1} and there is a sustained interest
in understanding their properties.
Thus, recently DyB$_2$C$_2$, \cite{Dy1} DyB$_6$, \cite{Dy2} 
UCu$_2$Sn, \cite{UCu2Sn} PrPb$_3$, \cite{PrPb3} YbAs, \cite{YbAs} 
YbSb \cite{YbSb}
were reported to undergo a quadrupolar ordering.

\section {The model} 
\label{sec:rd} 

As follows from electronic
band structure calculations of $\gamma$-Ce there exist
three conduction electrons per atom which form the $(6s6p5d)^3$ metallic band
and one localized 4$f$ electron. \cite{Ce_band,Sva,Jar,Laeg} 
In our previous work, Ref.\ \onlinecite{NM2}, we have already considered
electric multipole interactions between conduction electrons
and the localized 4$f$ electrons. 
Below we focus on the on- site and inter- site
correlations in the system and will simplify the model.
We consider the instantaneous
configuration $6s^25d4f$ as having the
largest statistical weight on a cerium site 
(in comparison with other possibilities 
such as $6s6p5d4f$, $6s^25d^2$, $6s^24f^2$ and etc.).
The two $6s$ electrons give only a spherically symmetrical
density on the cerium site and their lowest energy state
corresponds to a singlet. Therefore, 
as a first approximation we discard them as giving a closed shell. 
(Indeed, the level structure of $6s^25d4f$ corresponding to atomic
Ce~I is similar to that of $5d4f$ of La~II \cite{at_sp})
In fact, in doing so we omit the
$s-d$ electron transitions which can contribute to the quadrupolar
density. \cite{NM2} 
We are left then with one 5$d$ conduction electron and
one localized 4$f$ electron.
In the electronic band structure calculations the charge density
of the 5$d$ electron on the cerium site is considered as an average over all
occupied $\vec{k}$ states ($E(\vec{k},\alpha) \leq E_F$,
where $\alpha$ is the band index and $E_F$ is the Fermi level).
We have shown in Appendix~B of Ref.\ \onlinecite{NM1} that the electron
density is mainly spherical which corresponds to the standard
``muffin-tin" (MT) treatment in electronic band structure calculations.
The spherical density of 5$d$
and 4$f$ electrons will be the starting point in this work.
We consider the 5$d$ electron on a cerium center being instantaneously
coupled with the 4$f$ electron
and include in the model all corresponding intra- site interactions,
crystal electric field effects and inter- site multipolar
electric interactions.
From the technical point of view, this $fd$-model is a many
electron generalization of the concepts of Ref.\ \onlinecite{NM1}. 
We are aware that the model based on the $fd$-configuration 
is incomplete, but it has an advantage of taking into
account all intra- site interactions (often
referred to as Hund's rules) which are 
usually omitted in the electron band
structure calculations. \cite{LAPW,LMTO} Later we will briefly
discuss a possibility to refine our model with the help of
the valence bond (or Heitler-London) theory of chemical bonding. \cite{HL,Tin} 

We consider a face centered cubic (fcc) crystal of $N$ Ce atoms.
Each atomic site possesses one 4$f$ and one 5$d$ electron.
The position vector of an electron near a crystal lattice site $\vec{n}$
is given by
\begin{eqnarray}
 \vec{R}(\vec{n}) = \vec{X}(\vec{n})+\vec{r}(\vec{n}) .
\label{2.1} 
\end{eqnarray}
Here $\vec{X}(\vec{n})$ is the lattice vector which specifies
the centers of the atoms (or Ce-nuclei) on a rigid fcc lattice.
The radius vector $\vec{r}(\vec{n})$ is given in polar coordinates
by $(r(\vec{n}),\Omega(\vec{n}))$, where $r$ is the length and
$\Omega=(\Theta,\phi)$ stands for the polar angles.
We label the two-electron basis ket-vectors at a lattice site $\vec{n}$ 
by a single index $I_{fd}$ or, alternatively, by the pair of 
single electron indices $(i_f,i_d)$:
\begin{eqnarray}
   | I_{fd} \rangle_{\vec{n}} = 
   | i_f; \, i_d \rangle_{\vec{n}} .
\label{2.2} 
\end{eqnarray}
The index $i$ stands for the electron orbital and spin projection
quantum numbers.
The corresponding basis wave functions are  
\begin{eqnarray}
  \langle \vec{r}, \vec{r}\,'  | I_{fd} \rangle_{\vec{n}} = 
  \langle \vec{r} | i_f \rangle_{\vec{n}} \cdot
  \langle \vec{r}\,' | i_d \rangle_{\vec{n}} ,
\label{2.3} 
\end{eqnarray}
where
\begin{mathletters}
\begin{eqnarray}
   & &\langle \vec{r} | i_f \rangle_{\vec{n}} = 
   {\cal R}_f(r(\vec{n})) \langle \hat{n} | i_f \rangle , \label{2.4a} \\ 
   & &\langle \vec{r}\,' | i_d \rangle_{\vec{n}} = 
   {\cal R}_d(r'(\vec{n})) \langle \hat{n}' | i_d \rangle . \label{2.4b}  
\end{eqnarray}
\end{mathletters}
Here ${\cal R}_f$ and ${\cal R}_d$ are radial components of the 4$f$ and 
the 5$d$ electron, respectively; 
$\hat{n}$ stands for $\Omega(\vec{n})$. 
There are 14 orientational vectors (or spin-orbitals)
$\langle \hat{n} | i_f \rangle$ 
for a 4$f$ electron ($i_f$=1-14)
and 10 orientational vectors $\langle \hat{n} | i_d \rangle$ 
for a 5$d$ electron ($i_d$=1-10). These spin-orbitals can be written as
\begin{mathletters}
\begin{eqnarray}
 & &\langle \hat{n} | i_f \rangle=\langle \hat{n} | m_f \rangle\, u_s(s_z(f)) , 
 \label{2.5a} \\ 
 & &\langle \hat{n} | i_d \rangle=\langle \hat{n} | m_d \rangle \, u_s(s_z(d)) . 
 \label{2.5b}  
\end{eqnarray}
\end{mathletters}
Here $u_s$ is the spin function ($s=\pm$) for the spin projections
$s_z=\pm1/2$ on the $z$-axis. The orbital parts,
$\langle \hat{n} | m_f \rangle$ ($m_f$=1-7) and 
$\langle \hat{n} | m_d \rangle$ ($m_d$=1-5),
are expressed in terms of spherical harmonics 
$Y_l^m(\Omega)=\langle \hat{n}| l,m \rangle$. 
We find it convenient
to work with real spherical harmonics. We consider 
\begin{mathletters}
\begin{eqnarray}
   \left \{
   Y_3^0,\; Y_3^{1c},\; Y_3^{1s},\; Y_3^{2c}, \; Y_3^{2s},\;
   Y_3^{3c},\; Y_3^{3s} \right \} = \langle \hat{n}| m_f \rangle
\label{2.6a} 
\end{eqnarray}
for 4$f$ electron (corresponding to $m_f=1-7$)
and 
\begin{eqnarray}
  \left \{ Y_2^0,\; Y_2^{1c},\; Y_2^{1s},\; Y_2^{2c}, \; Y_2^{2s}
  \right \} = \langle \hat{n}| m_d \rangle
\label{2.6b} 
\end{eqnarray}
\end{mathletters}
for 5$d$ electron (corresponding to $m_d=1-5$).
We use the definition of real spherical harmonics of Ref.\ \onlinecite{Bra}
(see also explicit expressions (2.1) in Ref.\ \onlinecite{NM1}) that is
different from the definition of Condon and Shortley. \cite{CS}
The advantage of using the basis with real spherical harmonics is that the
matrix elements of Coulomb and exchange interactions stay real. 

The order of indices in (\ref{2.2}) and (\ref{2.3}) is important 
if we associate the first electron with the $f$ state $i_f$ while
the second with the $d$ state $i_d$. 
Then in addition to the vectors (\ref{2.2})
we have to consider the states described by the vectors 
$| i_d;\, i_f \rangle_{\vec{n}}$ (the first electron is in the $i_d$
state and the second is in the $i_f$ state). 
However, from the dynamical
equivalence of the electrons we can permute the spin-orbitals
to the standard order, Eq.\ (\ref{2.2}), by using
\begin{eqnarray}
   | i_d; \, i_f \rangle_{\vec{n}} 
   = -| i_f; \, i_d \rangle_{\vec{n}} ,
\label{2.7} 
\end{eqnarray}
since it requires the interchange of the two electrons.
In order to describe the same quantum state $(i_f;i_d)$ we will use
the basis vectors (\ref{2.2}) and apply (\ref{2.7}) when 
needed.
(Alternatively, one can use the procedure of antisymmetrization
of the basis vectors (\ref{2.2}) as described elsewhere.) 
Thus, our basis (\ref{2.2}) consists of 140 different vectors 
$| I_{fd} \rangle$. 

The ground state energy of the $(4f5d)$ electron system, $E_0$,
can be calculated in local density approximation (LDA)
with spherically symmetric Coulomb and exchange potentials.
Going beyond this model in atomic cerium, one has to take into
account multipolar on-site (also called intra- site) Coulomb interactions
and spin-orbit coupling.
In solid cerium, the interactions with conduction electrons and
inter- site Coulomb interactions still have to be added.

In the following we will study these effects within a unified formalism
based on a multipole expansion of the Coulomb potential and of the
systematic use of site symmetry of the crystal lattice.
For the case of on-site Coulomb interactions between two
electrons (charge $e=-1$) we have
\begin{eqnarray}
  V(\vec{R}(\vec{n}),\vec{R}'(\vec{n}))=
  {\frac{1}{|\vec{r}(\vec{n})-\vec{r}\,'(\vec{n})|}}.
\label{2.8} 
\end{eqnarray}
The multipole expansion in terms of site symmetry adapted functions (SAF's)
$S_{\Lambda}(\hat{n})$ reads:
\begin{mathletters}
\begin{eqnarray}
  V(\vec{R}(\vec{n}),\vec{R}'(\vec{n}'))=
  \sum_{\Lambda} 
  v_{\Lambda \Lambda}(r,r')\,
  S_{\Lambda}(\hat{n})\, S_{\Lambda}(\hat{n}'),  
\label{2.9a} 
\end{eqnarray}
where
\begin{eqnarray}
  v_{\Lambda \Lambda'}(r,r')\,
  = \left( {\frac {r^l_< }{r^{(l+1)}_> }} \right)
  {\frac {4\pi}{2l+1}} 
  \delta_{\Lambda \Lambda'} , 
\label{2.9b} 
\end{eqnarray}
\end{mathletters}
with $r_>=max(r,r')$, $r_<=min(r,r')$ and 
$\delta_{\Lambda \Lambda'}=\delta_{\tau \tau'} \delta_{l l'}$.
Clearly, the last expression is site independent.
The SAF's are linear combinations of spherical
harmonics and transform as irreducible representations
of the site point group, Ref.\ \onlinecite{Bra}.
The index $\Lambda$ stands for $(l,\tau)$, with $\tau=(\Gamma,\mu,k)$.
Here $l$ accounts for the angular dependence of the
multipolar expansion, $\Gamma$ denotes an
irreducible representation (in the present case of the 
group $O_h$),
$\mu$ labels the representations that occur more than once and
$k$ denotes the rows of a given representation.

On the other hand, the Coulomb interaction between two electrons at
different sites $\vec{n} \neq \vec{n}'$ (inter- site) reads
\begin{eqnarray}
  V(\vec{R}(\vec{n}),\vec{R}'(\vec{n}'))=
  {\frac{1}{|\vec{R}(\vec{n})-\vec{R}'(\vec{n}')|}}.
\label{2.10} 
\end{eqnarray}
The multipole expansion is given by
\begin{mathletters}
\begin{eqnarray}
 & & V(\vec{R}(\vec{n}),\vec{R}'(\vec{n}'))=
  \sum_{\Lambda \Lambda'} 
  v_{\Lambda \Lambda'}(\vec{n},\vec{n}';\,r,r')\,
  S_{\Lambda}(\hat{n})\, S_{\Lambda'}(\hat{n}'), \nonumber \\
 & & \label{2.11a} 
\end{eqnarray}
where
\begin{eqnarray}
 & & v_{\Lambda \Lambda'}(\vec{n},\vec{n}';\,r,r')\,
   =  \int \! d\Omega(\vec{n}) \int \! d\Omega'(\vec{n}')\,
      {\frac{  S_{\Lambda}(\hat{n})\, S_{\Lambda'}(\hat{n}')}
     {|\vec{R}(\vec{n})-\vec{R}'(\vec{n}')|}} . \nonumber   \\
 & & \label{2.11b} 
\end{eqnarray}
\end{mathletters}
The inter- site multipole expansion (\ref{2.11a}) 
is anisotropic and converges fast since \cite{Hei} 
\begin{eqnarray}
 v_{\Lambda \Lambda'}(\vec{n},\vec{n}';\,r,r') 
 \sim \frac{(r)^l (r')^{l'}}{|\vec{X}(\vec{n})-\vec{X}(\vec{n}')|^{l+l'+1}} .
 \label{2.12} 
\end{eqnarray}
Therefore, it is sufficient to consider
it only for nearest neighbors. From the practical point of view,
one can calculate 
$v_{\Lambda \Lambda'}(\vec{n},\vec{n}';\,r_0,r'_0)$ only for 
two fixed radii $r_0$ and $r'_0$. Then one obtains 
$v_{\Lambda \Lambda'}(\vec{n},\vec{n}';\,r,r')$ as a function
of $r$ and $r'$ by employing the dependence (\ref{2.12}).

\section {The intra- site interactions} 
\label{sec:cf} 

The interactions which we analyze in this section are
present already in atomic cerium. \cite{CS,at_sp}
We have considered a part of these correlations in 
previous work, \cite{NM2} and here we study all of them
in the framework of the $fd-$ model.
In fact, these multipole interactions 
are responsible for the electronic terms of atoms. \cite{CS}
We will see that their combined effect lowers
the energy of the cerium atom by $\sim 1-2$ eV in comparison
with the spherically symmetric case, yet usually they are not
taken into account in the electronic band structure
calculations in solids. 
Although our consideration in this section is based on the
original technique for multipole interactions, \cite{NM1} it
overlaps largely with the method of Condon and Shortley. \cite{CS}
However, for two reasons we have decided to briefly review it here. 
Firstly, we consider below a more general case which is
not limited by the $LS$ (Russell-Saunders) coupling and
the consideration of diagonal matrix elements. 
Secondly,
the results of this section are used to
describe crystal electric field effects and the phase
transition to the $Pa{\bar 3}$ structure. 

The direct matrix elements for the intra- site Coulomb
interactions are obtained if we consider
only the $f-f$ transitions for the first electron
and the $d-d$ transitions for the second.
We start from Eq.~(\ref{2.9a}) and obtain
\begin{eqnarray}
& &  \langle I_{fd} |_{\vec{n}}
  V(\vec{R}(\vec{n}),\vec{R}'(\vec{n}))
  | J_{fd} \rangle_{\vec{n}}^{Coul} 
 \nonumber \\
 & &  
  =\sum_{\Lambda} 
  v_{\Lambda \Lambda}^{F D}\,
  c_{\Lambda}(i_f j_f)\, c_{\Lambda}(i_d j_d), 
\label{3.1} 
\end{eqnarray}
where
\begin{eqnarray}
  v_{\Lambda \Lambda}^{F D} =
  \int \! dr\, r^2 \int \! dr'\, {r'}^2\,  
  {\cal R}_f^2(r)\, {\cal R}_d^2(r')\,
  v_{\Lambda \Lambda}(r,r')  
\label{3.2} 
\end{eqnarray}
accounts for the average radial dependence 
while $v_{\Lambda \Lambda}(r,r')$ is given by Eq.~(\ref{2.9b}).
We use the superscripts $F$ and $D$ in order to indicate that
we have transitions between two 4$f$ states ($F \equiv (f,f)$)
and the transitions between two 5$d$ states ($D \equiv (d,d)$).
The elements $c_{\Lambda}$ are defined by
\begin{mathletters}
\begin{eqnarray}
  c_{\Lambda}(i_f j_f) =  
  \int \! d\Omega \, \langle i_f |\hat{n}\rangle  
  S_{\Lambda}(\hat{n}) \langle \hat{n}|j_f \rangle ,
\label{3.3a} 
\end{eqnarray}
\begin{eqnarray}
  c_{\Lambda}(i_d j_d) =  
  \int \! d\Omega \, \langle i_d |\hat{n}\rangle  
  S_{\Lambda}(\hat{n}) \langle \hat{n}|j_d \rangle .
\label{3.3b} 
\end{eqnarray}
\end{mathletters}

The other possibility is to consider the transitions
$4f \rightarrow 5d$ for the first electron and the
transitions $5d \rightarrow 4f$ for the second.
This gives the exchange interactions and then we should
use (\ref{2.7}) in order to return to the standard order
of the spin-orbitals. We find
\begin{eqnarray}
& &  \langle I_{fd} |_{\vec{n}}
  V(\vec{R}(\vec{n}),\vec{R}'(\vec{n}))
  | J_{fd} \rangle_{\vec{n}}^{exch} 
 \nonumber \\
 & &  
  =-\sum_{\Lambda} 
  v_{\Lambda \Lambda}^{(fd) (df)}\,
  c_{\Lambda}(i_f j_d)\, c_{\Lambda}(i_d j_f) , 
\label{3.4} 
\end{eqnarray}
where
\begin{eqnarray}
  v_{\Lambda}^{(fd)}{}_{\Lambda}^{(df)} &=&
  \int \! dr\, r^2 \int \! dr'\, {r'}^2\,  
  {\cal R}_f(r){\cal R}_d(r)\, {\cal R}_d(r'){\cal R}_f(r')\,
  \nonumber \\
  & & \times v_{\Lambda \Lambda}(r,r') ,  
\label{3.5} 
\end{eqnarray}
and 
\begin{mathletters}
\begin{eqnarray}
  c_{\Lambda}(i_f j_d) =  
  \int \! d\Omega \, \langle i_f |\hat{n}\rangle  
  S_{\Lambda}(\hat{n}) \langle \hat{n}|j_d \rangle ,
\label{3.6a} 
\end{eqnarray}
\begin{eqnarray}
  c_{\Lambda}^{df}(i_d j_f) =  
  \int \! d\Omega \, \langle i_d |\hat{n}\rangle  
  S_{\Lambda}(\hat{n}) \langle \hat{n}|j_f \rangle .
\label{3.6b} 
\end{eqnarray}
\end{mathletters}
We observe that in the basis with real orbitals,
Eq.\ (\ref{2.6a},b), and with real functions $S_{\Lambda}$
the coefficients $c_{\Lambda}$ are real and we get
\begin{eqnarray}
  c_{\Lambda}(i_d j_f) =  
  c_{\Lambda}(j_f i_d) .
\label{3.7} 
\end{eqnarray}

We start with the description of spherically symmetric terms ($l=0$)
corresponding to the trivial function $S_0=1/\sqrt{4\pi}$.
The coefficients $c_{\Lambda}$ in (\ref{3.3a},b) become
diagonal,
\begin{eqnarray}
  c_{l=0}(i_d j_d) = \frac{1}{\sqrt{4\pi}} \delta_{i_d j_d}, \;\;  
  c_{l=0}(i_f j_f) = \frac{1}{\sqrt{4\pi}} \delta_{i_f j_f} ,
\label{3.8} 
\end{eqnarray}
while $c_{l=0}(i_f j_d)=c_{l=0}(i_d j_f)=0$. 
Hence, we obtain a contribution to $\langle I |V| J \rangle^{Coul}$
which is proportional to the unit matrix.
Since it corresponds only to a shift of the ground state energy,
it is irrelevant.

In considering the other
contributions (with $l > 0$) in (\ref{3.1}) and (\ref{3.4}) 
we will take advantage of the selection rules
imposed by the coefficients $c_{\Lambda}$, Eqs. (\ref{3.3a},b)
and (\ref{3.6a},b). First of all, we notice that
the coefficients $c_{\Lambda}$ are diagonal in terms
of spin components $u_s$. 
From the theory of addition of angular momenta
(see, for example, Ref.\ \onlinecite{Tin})
we know that nonzero coefficients $c_{\Lambda}(i_f j_f)$
can occur if $l$ (in $\Lambda$) equals to 0,1,2,...,6, Eq.\ (\ref{3.3a}).
Furthermore, the odd values of $l$ are excluded due to
the parity of the integrand in (\ref{3.3a}) and finally we obtain
that $l=0$, 2, 4 and 6.
Analogously, for the $d-d$ transitions the allowed
coefficients are with $l=0$, 2 and 4, Eq.\ (\ref{3.3b}). For the 
$f-d$ transitions, Eq.\ (\ref{3.6a},b) we find
that $l=1$, 3 and 5. 
Next we notice that if the radial parts, ${\cal R}_f$, ${\cal R}_d$,
are the same for all spin-orbitals of 4$f$ and 5$d$ states,
correspondingly, then the integrals (\ref{3.2}) and (\ref{3.5})
depend only on $l$. We can condense the notation,
$v_l^{F D}=v_{\Lambda \Lambda}^{F D}$ for $l=0$, 2 and 4;
and $v_{l'}^{(fd)}=v_{\Lambda}^{(fd)}{}_{\Lambda}^{(df)}$ for
$l'=1$, 3 and 5. In fact, these integrals are proportional
to $F^k$ and $G^k$ in the notations of Condon and Shortley, \cite{CS}
\begin{eqnarray}
  & & v_2^{F D}=\frac{4\pi}{5} F^2(4f,5d), \;\;\;
  v_4^{F D}=\frac{4\pi}{9} F^4(4f,5d)  \nonumber \\
  & & v_1^{(fd)}=\frac{4\pi}{3} G^1(4f,5d), \;\;\;
  v_3^{(fd)}=\frac{4\pi}{7} G^3(4f,5d), \label{3.9} \\
  & & v_5^{(fd)}=\frac{4\pi}{11} G^5(4f,5d) .  \nonumber  
\end{eqnarray}
(Notice however that our coefficients $c_{\Lambda}$ are
different from those of Condon and Shortley. \cite{CS})
We have estimated the integrals in (\ref{3.9}) from the
radial dependences ${\cal R}_f$ and ${\cal R}_d$ obtained 
from calculations on a 
cerium atom in local density approximation (LDA).
We have found that $v_2^{F D}=85858$, $v_4^{F D}=24260$,
$v_1^{(fd)}=83128$, $v_3^{(fd)}=26563$ and $v_5^{(fd)}=12584$,
in K (Kelvin).

Finally, we rewrite expressions (\ref{3.1}) and (\ref{3.4}) in
matrix form as
\begin{mathletters}
\begin{eqnarray}
 & & \langle I_{fd} |
  V(intra)
  | J_{fd} \rangle^{Coul} = \nonumber \\ 
  & & 
 v_2^{F D} c_2^{F D}(I_{fd}| J_{fd}) + v_4^{F D} c_4^{F D}(I_{fd}| J_{fd}) ,
\label{3.10a} 
\end{eqnarray}
and
\begin{eqnarray}
 & & \langle I_{fd} |
  V(intra)
  | J_{fd} \rangle^{exch} = -[ 
 v_1^{(fd)} c_1^{(fd)}(I_{fd}| J_{fd}) \nonumber \\  
 & & + v_3^{(fd)} c_3^{(fd)}(I_{fd}| J_{fd})
 + v_5^{(fd)} c_5^{(fd)}(I_{fd}| J_{fd})\, ] ,
\label{3.10b} 
\end{eqnarray}
\end{mathletters}
Here the direct Coulomb matrices $c_l^{F D}(I_{fd}| J_{fd})$ are defined as
\begin{mathletters}
\begin{eqnarray}
 c_l^{F D}(I_{fd}| J_{fd}) = \left( \sum_{\tau} 
 c_{(l,\tau)}(i_f j_f)\,c_{(l,\tau)}(i_d j_d) \right) ,
\label{3.11a} 
\end{eqnarray}
where $l=2$ and 4; and the three ``exchange" matrices 
$c_l^{(fd)}(I_{fd}| J_{fd})$ ($l=1$, 3 and 5) are given by
\begin{eqnarray}
 c_l^{(fd)}(I_{fd}| J_{fd}) = \left( \sum_{\tau} 
 c_{(l,\tau)}(i_f j_d)\,c_{(l,\tau)}(i_d j_f) \right) .
\label{3.11b} 
\end{eqnarray}
\end{mathletters}
We solve the secular problem for the $140 \times 140$ matrix of intra- site
interactions
\begin{eqnarray}
 & &  \langle I_{fd} | V(intra) | J_{fd} \rangle =  \nonumber   \\
 & &  \langle I_{fd} | V(intra) | J_{fd} \rangle^{Coul} +
  \langle I_{fd} | V(intra) | J_{fd} \rangle^{exch} ,
\label{3.12} 
\end{eqnarray}
and obtain the ten energy levels $E_{fd}$ quoted in 
column 2 and 3 of Table I.
This term spectrum corresponds to the usual $LS$ (Russell-Saunders)
coupling without spin-orbit interactions.
We have also checked our results by working with the eigenvectors
of (\ref{3.12}) which can be obtained independently
by exploiting the formulas (Table $4^3$ of Ref.\ \onlinecite{CS}) of vector
addition of angular momenta ($j_1=3$ and $j_2=2$).  
The levels of Table I correspond to the following parameters
in the notations of Condon and Shortley, \cite{CS}
$F_2=325.4$, $F_4=25.1$, $G_1=567$, $G_3=47$ and $G_5=7.2$ (in K)
(These parameters should not be confused with those in (\ref{3.9})).
%
\begin{table} 
\caption{
Term energies of the $fd$ configuration in the absence of
spin-orbit coupling,  $E_m$ and $\triangle E$ 
stand for the singlet-triplet energy
means and the singlet-triplet energy differences, respectively.
$L$ is the orbital quantum number of the two electrons.
The energy corresponding to the spherically symmetric
description is taken as zero.
All energies are in units K.
\label{tab1}     } 
 
 \begin{tabular}{c c c c c} 
 $L$ & singlets & triplets & $E_m$ & $\triangle E$ \\
\tableline 
 $P$ & 13541.8 &  5384.5 &  9463.1 &  8157.3 \\
 $D$ & -3011.8 &  1951.8 & -530.0  & -4963.5 \\
 $F$ &  2767.8 & -6616.1 & -1924.1 &  9383.9 \\
 $G$ & -9378.4 & -1485.3 & -5431.8 & -7893.1 \\
 $H$ & 12310.8 & -5653.4 &  3328.7 & 17964.2 
 \end{tabular} 
\end{table} 

We now consider the effect of the spin-orbit coupling.
Starting with the spherically symmetric LDA 
calculation of a cerium atom we
obtain that in the one-electron approximation
$\triangle_{so}(f)=E_f(7/2)-E_f(5/2)=4003.4$ K
and $\triangle_{so}(d)=E_d(5/2)-E_d(3/2)=2344.0$ K.
This gives for the spin-orbit coupling constants
$\zeta_f=1143.8$ K and $\zeta_d=937.6$ K.
Therefore, a typical value of spin-orbit splitting
is $\sim 1000$ K which shows that it can not be treated
as a small perturbation to the $LS$ term scheme, Table I.
In order to take into account the spin-orbit coupling exactly
we have to consider the operator 
\begin{eqnarray}
 V_{so}=V_{so}(f)+V_{so}(d) ,
\label{3.12'} 
\end{eqnarray}
where
\begin{mathletters}
\begin{eqnarray}
 & & V_{so}(f)=\zeta_f \, \vec{L}(f) \cdot \vec{S}(f) ,
\label{3.14a}  \\
 & & V_{so}(d)=\zeta_d \, \vec{L}(d) \cdot \vec{S}(d) .
\label{3.14b}
\end{eqnarray}
\end{mathletters}
The matrix of interactions reads
\begin{eqnarray}
 & & \langle I_{fd} | V_{so} | J_{fd} \rangle = \left(
 \langle i_f | V_{so}(f) | j_f \rangle\, \delta_{i_d j_d} 
 + \langle i_d | V_{so}(d) | j_d \rangle\, \delta_{i_f j_f} \right) .
 \nonumber  \\
 & & \label{3.15} 
\end{eqnarray}
Since we know the matrix elements $\langle i_f | V_{so}(f) | j_f \rangle$
and $\langle i_d | V_{so}(d) | j_d \rangle$ (see for example the explicit
Eq.\ (A.2) of Ref.\ \onlinecite{NM1}) we can calculate the matrix elements
for the $fd$-configuration.
We now solve the secular problem for the sum of the
intra-site and spin-orbit interactions, starting from the matrix of
\begin{eqnarray}
  \left. U \right|_{intra} = V(intra) + V_{so} ,
\label{3.16} 
\end{eqnarray}
and obtain 20 energy levels $\{ E_{fd} \}$. 
Since we are interested only in the
lowest levels, we quote in Table II the first six out of the 20 levels.
%
\begin{table} 
\caption{
Calculated lowest term energies of the $fd$ configuration with the
spin-orbit coupling and the Lande $g$-factors (columns 2-4),
$\triangle E=E_{fd}-E_{fd}(^1 G_4)$. 
Last two columns are experimental data,
Ref.~34.
All energies are in units K.
\label{tab2}     } 
 
 \begin{tabular}{c c c c | c c} 
   &  $E_{fd}$ & $g$ & $\triangle E$  & Ce I & La II $(fd)$ \\
\tableline
 $^1 G_4$ &  -10570.3 & 0.9323 & 0      & 0      & 0 \\
 $^3 F_2$ &   -9226.5 & 0.7030 & 1343.8 & 329.2  & 881.7 \\
 $^3 H_4$ &   -8155.1 & 0.8958 & 2415.2 & 1840.9 & 1764.7 \\
 $^3 F_3$ &   -7048.8 & 1.0825 & 3521.5 & 2393.0 & 2354.5 \\
 $^3 H_5$ &   -6449.4 & 1.0334 & 4120.9 & 3177.9 & 2850.7 \\
 $^3 G_3$ &   -4247.3 & 0.7659 & 6323.0 & 1998.5 & 5472.8 
 \end{tabular} 
\end{table}
(The procedure of calculation of magnetic moments and the Lande $g$-factors
is given in Appendix~A.) 

Comparison with the experimental data \cite{at_sp} on a Ce atom shows that 
the order of the three lowest levels, $^1 G_4$, $^3 F_2$
and $^3 H_4$, with the multiplet $^1 G_4$ as the ground state is correct.
On the other hand, the experimental data show that 
in atomic Ce different levels
(such as $^1 G$, $^3 F$, $^3 H$) are considerably mixed up.
Due to the strong spin-orbit coupling there is an
appreciable mixing between $^1 G_4$ and $^3 H_4$,
and between $^3 F_2$ and $^1 D_2$. The ground state of a Ce atom
has 55\% of $^1 G_4$ and 29\% of $^3 H_4$. \cite{at_sp} The next level which
lies only 329 K above the ground state has 66\% of $^3 F_2$
and 24\% of $^1 D_2$. \cite{at_sp} 
In conclusion, although
the actual atomic spectra of cerium differ somewhat from our results 
obtained
for the $fd-$ configuration, Table II, our approach captures
the main properties and we will use it for Ce in the solid state.
In the following we will extend the calculations of the energy
level scheme by including inter-site interactions.
We will treat separately the $\gamma$ and the $\alpha$ phase
of solid Ce.

\section {Inter- site interactions} 
\label{sec:qo}

In this section we will first consider the matrix of inter-site interactions 
for the $(fd)$ system on a crystal in general.
Next we will show that the interaction is largely simplified by crystal symmetry 
and derive the crystal electric field (CEF).
We calculate the energy spectrum of the $(fd)$ system in presence of
the crystal electric field in the disordered cubic phase.

We start from expression (\ref{2.11a}) and write it in the space 
of two electron state vectors $| I_{fd} \rangle$.
Carrying out the angular integrations $d\Omega(\vec{n})$,
$d\Omega'(\vec{n})$, $d\Omega(\vec{n}')$, $d\Omega'(\vec{n}')$,
we obtain
\begin{eqnarray}
& & \langle I_{fd} |_{\vec{n}} \langle I'_{fd} |_{\vec{n}'}
  V(\vec{R}(\vec{n}),\vec{R}'(\vec{n}'))
  |J'_{fd} \rangle _{\vec{n}'}| J_{fd} \rangle _{\vec{n}} = 
     \nonumber \\
& & \sum_{\alpha \beta} \sum_{\alpha' \beta'} \sum_{\Lambda \Lambda'}
  v_{\Lambda}^{\alpha \alpha}{\,}_{\Lambda'}^{\alpha' \alpha'}(\vec{n}-\vec{n}')
     \nonumber \\
& & \times 
  \left\{ c_{\Lambda}(i_{\alpha} j_{\alpha})\,\delta(i_{\beta} j_{\beta}) \right\} 
  \left\{ c_{\Lambda'}(i_{\alpha'} j_{\alpha'})\, \delta(i_{\beta'} j_{\beta'})  
  \right\} ,  
\label{4.1} 
\end{eqnarray}
Here each of the indices $\alpha$, $\alpha'$, $\beta$, $\beta'$ runs
over the labels $f$, $d$.
The coefficients $c_{\Lambda}$ are defined by Eqs. (\ref{3.6a}) and (\ref{3.6b}),
while $\delta(i_{\beta} j_{\beta})$ stands for the Kronecker delta symbol.
For $\alpha=f$, we have $\beta=d$ and for $\alpha=d$, $\beta=f$, with a
similar correspondence between $\alpha'$ and $\beta'$.
The inter- site interaction element 
$v_{\Lambda}^{\alpha \alpha}{\,}_{\Lambda'}^{\alpha' \alpha'}$
is given by
\begin{eqnarray}
  v_{\Lambda}^{\alpha \alpha}{\,}_{\Lambda'}^{\alpha' \alpha'}(\vec{n}-\vec{n}') &=&
  \int \! dr\, r^2 \int \! dr'\, {r'}^2\,  \nonumber \\
  & \times & {\cal R}_{\alpha}^2(r)\, {\cal R}_{\alpha'}^2(r')\,
  v_{\Lambda \Lambda'} (\vec{n},\vec{n}';\, r,r') ,
\label{4.2} 
\end{eqnarray}

Notice that only direct Coulomb interactions are present
in (\ref{4.1}), and with the help of the selection rules for
the coefficients $c_{\Lambda}$ we conclude that only
the interactions with $l=0$, 2, 4 and 6 have to be considered. 

\subsection{The crystal electric field ($\gamma$-phase)}

In the $\gamma$-phase the electronic density is compatible with
the crystal structure $Fm{\bar 3}m$.
At each atomic site the crystal electric field (CEF) has the
point group symmetry $O_h$.
In lowest approximation, the CEF corresponds to the potential
experienced by a charge at a central site $\vec{n}$, when
spherically symmetric ($l'=0$) contributions from charge
densities at the twelve neighboring sites $\vec{n}'$ on the
fcc lattice and similar terms from the electronic density in the
interstitial regions are taken into account.
Previously \cite{NM1,NM2} these effects were studied for a single 4$f$
electron per Ce atom. Here we present an extension to the ($4f5d$)
system.

In crystal field approximation the functions $S_{\Lambda'}(\vec{n}')$
at any of the twelve sites $\vec{n}'$ reduce to $Y_0^0=1/\sqrt{4\pi}$.
We will write an index $0$ for $\Lambda'$ $(l'=0,A_{1g})$.
The coefficients $c_{\Lambda'}$ in Eq.\ (\ref{4.1}) now reduce to
\begin{eqnarray}
 c_0(i_{\alpha'} j_{\alpha'}) = {\frac{1}{\sqrt{4\pi}}}
 \delta(i_{\alpha'},j_{\alpha'}) .  
\label{4.3} 
\end{eqnarray}
At the central site $\vec{n}$, the electronic density has full cubic symmetry.
We denote the corresponding SAF's by $S_{\Lambda_1}(\vec{n})$, 
$\Lambda_1 \equiv (l,A_{1g})$, 
where $A_{1g}$ stands for the unit representation
of the cubic site group $O_h$. We retain the functions for $l=4$ and
$l=6$, which correspond to the cubic harmonics $K_4$ and $K_6$.
The selection rules imply that the $d-d$ transitions are perturbed by
$K_4$ only, while for the $f-f$ transitions 
both $K_4$ and $K_6$ are relevant.
Expression (\ref{2.11a}) reduces to
\begin{mathletters}
\begin{eqnarray}
 & & V(\vec{R}(\vec{n}),\vec{R}'(\vec{n}'))=\frac{1}{\sqrt{4\pi}}
 \sum_{\Lambda_1} v_{\Lambda_1 0}(\vec{n},\vec{n}';\, r,r')\,
 S_{\Lambda_1}(\vec{n}) . \nonumber \\
 & & \label{4.4a}
\end{eqnarray}
The elements
\begin{eqnarray}
  & & v_{\Lambda_1 0}(\vec{n},\vec{n}';\,r,r')\,
   =    \nonumber \\
  & &  \frac{1}{\sqrt{4\pi}}
   \int \! d\Omega(\vec{n}) \int \! d\Omega'(\vec{n}')\,
      {\frac{  S_{\Lambda_1}(\hat{n}) }
     {|\vec{R}(\vec{n})-\vec{R}'(\vec{n}')|}}   
\label{4.4b}
\end{eqnarray}
\end{mathletters}
have the same value for all twelve neighbors $\vec{n}'$ on the fcc lattice.
In addition they are independent of $r'$, as follows from expression (\ref{2.12})
for $l'=0$.
We then define the crystal field operator by
\begin{eqnarray}
 V_{CF}(\vec{R}(\vec{n}))=\frac{12}{\sqrt{4\pi}}
 \sum_{\Lambda_1} v_{\Lambda_1 0}(\vec{n},\vec{n}';\, r, {\not r'})\,
 S_{\Lambda_1}(\vec{n}) .
\label{4.5} 
\end{eqnarray}
Returning to expression (\ref{4.2}) we write within the crystal field
approximation
\begin{mathletters}
\begin{eqnarray}
 v_{\Lambda_1}^{\alpha \alpha}{\,}_0^{\alpha' \alpha'}(\vec{n}-\vec{n}')=
 v_{\Lambda_1}^{\alpha \alpha}{}_0^{\bullet} \cdot Q_{\alpha'},
\label{4.6a}
\end{eqnarray}
where
\begin{eqnarray}
  v_{\Lambda_1}^{\alpha \alpha}{}_0^{\bullet}  = 
  \int dr\,r^2\, {\cal R}_{\alpha}^2(r) \,
  v_{\Lambda_1\, 0}(\vec{n},\vec{n}';\, r,{\not  r'})
\label{4.6b}
\end{eqnarray}
and
\begin{eqnarray}
  Q_{\alpha'}=\int dr'\,{r'}^2\,{\cal R}_{\alpha'}^2(r') . 
\label{4.6c}
\end{eqnarray}
\end{mathletters}
Notice that $Q_{\alpha'}$ stands for $Q_f$ or $Q_d$ which are charges
(in units $e$) of the 4$f$ or 5$d$ electron.
As before \cite{NM1,NM2} the integration is taken over $0<r'<R_{MT}$,
where $R_{MT}$ is the radius of the muffin-tin sphere.
Besides 4$f$ and 5$d$ electrons we can also consider similar
contributions from 6$s$ electrons and nuclei belonging
to nearest neighbors. Notice that the interaction
parameters $v_{\Lambda_1}^{f f}{}_{0}^{\bullet}$ and  
$v_{\Lambda_1}^{d d}{}_{0}^{\bullet}$, Eq.\ (\ref{4.6b}),
remain the same for all these contributions and all we have to do is
to collect the charges together. 
Finally, after summation over 12 nearest neighbors and simplifications,
we obtain
\begin{eqnarray}
 & & \langle I_{fd} |_{\vec{n}}  
  V_{CF}(\vec{R}(\vec{n})) | J_{fd} \rangle _{\vec{n}} = 
   \nonumber \\
& &    \sum_{\Lambda_1} 
\left[ B_{\Lambda_1}^f\, c_{\Lambda_1}(i_f j_f)\, \delta(i_d j_d) 
+  B_{\Lambda_1}^d\, c_{\Lambda_1}(i_d j_d)\, \delta(i_f j_f) \right] ,
\label{4.7}
\end{eqnarray}
where
\begin{mathletters}
\begin{eqnarray}
 & &  B_{\Lambda_1}^f = 
  {\frac {12}{\sqrt{4 \pi}}}\, Q_{eff}\,e\, v_{\Lambda_1}^{F}{\,}_0^{\bullet} , 
  \label{4.8a}  \\ 
 & & B_{\Lambda_1}^d = 
  {\frac {12}{\sqrt{4 \pi}}}\, Q_{eff}\,e\, v_{\Lambda_1}^{D}{\,}_0^{\bullet} .  
  \label{4.8b} 
\end{eqnarray}
\end{mathletters}
Here again we write $D$ for $(d d)$ and $F$ for $(f f)$. We take as an effective 
charge $Q_{eff}=Q_{MT}$, the total charge inside a MT-sphere.
In contradistinction to our previous work \cite{NM1,NM2} here
we have neglected the effect of interstitial charges.
Previously it was found that a {\em homogeneous} distribution of negative
charge in the interstices increases the effective charge by an amount
2.85$Q_{MT}$. This fact due to the angular dependence of the leading cubic harmonic
$S_{\Lambda_1}$, $\Lambda_1=\,(l=4,A_{1g})$ in Eq.\ (\ref{4.4b}).
Indeed $K_4(\hat{n})$ is positive and maximum along the cubic direction
$[100]$ (the centers of the interstices) and negative and small along
$[110]$ (the sites $\vec{n}'$). On the other hand, if we consider 
an inhomogeneous charge distribution where most of the electronic density 
in the interstices is located close to $[110]$, then the contribution
to the crystal field from interstitial charges can be assumed to be
negligibly small. We observe that previously the inclusion of contributions
from a homogeneous charge distribution in the interstices has led to
an overestimation of calculated crystal field splitting in comparison
with the experimental values. \cite{NM2}

In practice, it is convenient to calculate 
$v_{\Lambda_1}^{\alpha \alpha}{}_{0}^{\bullet}$ from
\begin{mathletters}
\begin{eqnarray}
  v_{\Lambda_1}^{\alpha \alpha}{}_{0}^{\bullet} =  
  \frac{v_{\Lambda_1\, 0}(\vec{n},\vec{n}'; R_{MT},{\not  r}')}{R_{MT}^l}
   q_l^{\alpha} ,
\label{4.9a}
\end{eqnarray}
where
\begin{eqnarray}
 q_l^{\alpha}= \int dr'\,{r'}^{\,(l+2)}\, {\cal R}_{\alpha}^2(r')  .
\label{4.9b}
\end{eqnarray}
\end{mathletters}
Then the CEF operator (\ref{4.5}), with an effective charge $Q_{eff}$,
can be rewritten as
\begin{mathletters}
\begin{eqnarray}
  V_{CF}(\vec{R}(\vec{n}))= \sum_{\Lambda_1} B_{\Lambda_1}\, 
  S_{\Lambda_1}(\hat{n}) \, r^l ,  
\label{4.10a}
\end{eqnarray}
where \cite{NM3}
\begin{eqnarray}
  B_{\Lambda_1}={\frac {12}{\sqrt{4 \pi}}}\, Q_{eff}\,e\,
  \frac{v_{\Lambda_1\, 0}(\vec{n},\vec{n}'; R_{MT},{\not  r}')}{R_{MT}^l} .
\label{4.10b}
\end{eqnarray}
\end{mathletters}

Our calculations for $\gamma$-Ce ($a$=9.753 a.u.) yield 
$Q_{MT}=+0.9136\,|e|$, \cite{IN}
$q_4^d/R_{MT}^4$=0.22271, $q_4^f/R_{MT}^4$=0.03604, 
$q_6^f/R_{MT}^6=0.01739$ (in a.u.), and
$B_4^d=$1198.4, $B_4^f=$193.9, $B_6^f$=74.4 (all in K). \cite{rem2} 
We then consider the Hamiltonian
\begin{eqnarray}
   H^{\gamma}(\vec{n})= \left. U \right|_{intra} + V_{CF}(\vec{n}) ,
\label{4.11}
\end{eqnarray}
which we associate with the $\gamma$ phase of Ce. 
By diagonalizing $H^{\gamma}$ we have found that
in the cubic CEF the twenty atomic- like levels of cerium
are split into 58 distinct sublevels which can be labeled by
single valued irreducible representations $A_1(\Gamma_1)$,
$E(\Gamma_3)$, $T_1(\Gamma_4)$ and $T_2(\Gamma_5)$
of $O_h$. In particular, 
three lowest levels of cerium are split 
according to the following scheme, \cite{Lea,Tin} 
\begin{mathletters}
\begin{eqnarray}
& & ^1 G_4 \rightarrow A_1 + T_1 + E + T_2 ,   \label{4.12a} \\
& & ^3 F_2 \rightarrow T_2 + E ,               \label{4.12b} \\
& & ^3 H_4 \rightarrow T_2 + A_1 + T_1 + E .   \label{4.12c} 
\end{eqnarray}
\end{mathletters}
The calculated splittings of these levels are quoted in Table III.
In presence of a magnetic field, there occurs an additional
splitting of the triplets. The corresponding magnetic moments are
given in the last column of Table III.
(Details of the calculations can be found in Appendix~A.)
In the second part of the present section we will study the energy levels
in the ordered phase.
%
\begin{table} 
\caption{
Lowest levels of the energy spectrum of  $H^{\gamma}=U|_{intra}+V_{CF}$,  
 $\gamma$-Ce. 
Numbers in parentheses stand for degeneracy;  
$\triangle \varepsilon_1=$1409.6~K, 
$\triangle \varepsilon_2=$2499.4~K;
the site group is $O_h$. 
\label{tab3}     } 
 
 \begin{tabular}{l c r c c } 
  $\Gamma,\;\mu$ &
 & $\varepsilon_i$, in K & $(\varepsilon_i-\varepsilon_1)$, in K & 
                                            ${\cal M}_z$, in $\mu_B$\\
\tableline
 $A_1$, 1 &(1) & -10661.1 &   0.0 &  0 \\
 $T_1$, 1 &(3) & -10613.7 &  47.4 & $\pm$0.4596; 0 \\
 $E$, 1   &(2) & -10580.9 &  80.2 &  0; 0 \\
 $T_2$, 1 &(3) & -10505.6 & 155.5 & $\pm$2.3112; 0 \\
\noalign{\smallskip}\hline\noalign{\smallskip} 
 $T_2$, 2 &(3) & -9251.5 &  $\triangle \varepsilon_1$      & $\pm$0.6634; 0 \\
 $E$, 2   &(2) & -9224.4 &  $\triangle \varepsilon_1$+27.1 & 0; 0 \\
\noalign{\smallskip}\hline\noalign{\smallskip} 
 $T_1$, 2 &(3) & -8161.7 &  $\triangle \varepsilon_2$      & $\pm$2.2447; 0 \\
 $T_2$, 3 &(3) & -8154.2 &  $\triangle \varepsilon_2$+7.5  & $\pm$0.4503; 0 \\
 $A_1$, 2 &(1) & -8145.9 &  $\triangle \varepsilon_2$+15.8  & 0 \\
 $E$, 3   &(2) & -8137.5 &  $\triangle \varepsilon_2$+24.2 & 0; 0  
 \end{tabular} 
\end{table}

%
%
\subsection{Quadrupolar ordering ($\alpha$-phase)}

In the cubic phase ($\gamma$-Ce) with the $Fm{\bar 3}m$ 
space symmetry the nontrivial electron density 
distribution is given by cubic harmonics $K_4(\Omega)$ and $K_6(\Omega)$
with $l=4$ and 6. All quadrupole densities with $l=l'=2$ 
average to zero.
Only fluctuations of electric quadrupoles are allowed 
in the interaction Eq.~(\ref{4.1}) and those lead
to an effective attractive interaction at the $X$ point of the
Brillouin zone. \cite{NM1,NM2}
This interaction drives a transition to a new phase which is 
characterized by an ordering of 
electric quadrupoles such that the space group symmetry is $Pa{\bar 3}$.
This order-disorder transition is accompanied by a contraction of the
crystal lattice which stays cubic.
We have associated this phase transition with the $\gamma \rightarrow \alpha$
transition of Ce.
In real space the $Pa{\bar 3}$ ordering implies 
the appearance of four distinct
sublattices of simple cubic structure (see Fig.\ 3 of Ref.\ \onlinecite{NM1}).
We label these sublattices which contain the sites (0,0,0)
$(a/2)$(0,1,1), $(a/2)$(1,0,1) and $(a/2)$(1,1,0) by $\{ \vec{n}_p \}$,
$p=1-4$, respectively.
In principle, one can proceed as in Ref.\ \onlinecite{NM1} and derive
an effective mean-field Hamiltonian.
Here we will start from the crystal in real space and consider
the following four quadrupolar SAF's corresponding to the
four sublattices of $Pa{\bar 3}$:
\begin{mathletters}
\begin{eqnarray}
 & & {\cal S}_{ \{n_1\} }(\Omega)=
 \frac{1}{\sqrt{3}}(S_1(\Omega)+S_2(\Omega)+S_3(\Omega)) ,   
  \label{4.12'a}  \\ 
 & & {\cal S}_{ \{n_2\} }(\Omega)=
 \frac{1}{\sqrt{3}}(-S_1(\Omega)-S_2(\Omega)+S_3(\Omega)) ,   
  \label{4.12'b}  \\ 
 & & {\cal S}_{ \{n_3\} }(\Omega)=
 \frac{1}{\sqrt{3}}(S_1(\Omega)-S_2(\Omega)-S_3(\Omega)) ,   
  \label{4.12'c}  \\ 
 & & {\cal S}_{ \{n_4\} }(\Omega)=
 \frac{1}{\sqrt{3}}(-S_1(\Omega)+S_2(\Omega)-S_3(\Omega)) .   
  \label{4.12'd}   
\end{eqnarray}
\end{mathletters}
Here we use the short notations $S_k \equiv S_{(\ell=2,T_{2g},k=1-3)}$
for real spherical harmonics $Y_2^{1s}$, $Y_2^{1c}$ and $Y_2^{2s}$
which belong to a three dimensional irreducible representation $T_{2g}$
of $O_h$.
(These spherical harmonics are proportional to the Cartesian components
$yz$, $zx$ and $xy$ for $k=1-3$.)

Below we consider the inter- site quadrupole interactions 
$V^{QQ}(\vec{n},\vec{n}')$  which involve
only the functions (\ref{4.12'a}-d). (There are also SAF's with
$l=4$ and 6 allowed by the $Pa{\bar 3}$ symmetry \cite{Bra} 
but those lead to weaker multipole interactions, Eq.\  (\ref{2.12}).)
We then rewrite Eq.\ (\ref{4.1}) for a case when $\vec{n} \in \{n_1\}$
and $\vec{n}' \in \{n_{p'}\}$ ($p'=$2, 3,~4):
\begin{eqnarray}
& & \langle I_{fd} |_{\vec{n}} \langle I'_{fd} |_{\vec{n}'}
  V^{QQ} (\vec{n}, \vec{n}')
  |J'_{fd} \rangle _{\vec{n}'}| J_{fd} \rangle _{\vec{n}} =       
  -\sum_{\alpha \alpha'} 
   \frac{\gamma^{\alpha \alpha \, \alpha' \alpha'} }{3} \, 
   \nonumber \\
 & &  \times
  \left\{ c_{ \{n_1\} }(i_{\alpha} j_{\alpha})\,\delta(i_{\beta} j_{\beta}) 
  \right\} 
  \left\{ c_{ \{n_{p'}\} }(i_{\alpha'} j_{\alpha'})\, \delta(i_{\beta'} j_{\beta'})  
  \right\} , \nonumber \\ 
\label{4.14}
\end{eqnarray}
where, as before, $\alpha,\alpha',\beta,\beta'$ run over $f$ and $d$,
and where the same exclusion rules (if $\alpha=f$ then $\beta=d$ and vice versa)
hold between $\alpha$ and $\beta$
as well as between $\alpha'$ and $\beta'$.
Here we have
\begin{eqnarray}
  \gamma^{\alpha \alpha \, \alpha' \alpha'} &=&
  \int \! dr\, r^2 \int \! dr'\, {r'}^2\,  \nonumber \\
  & \times & {\cal R}_{\alpha}^2(r)\, {\cal R}_{\alpha'}^2(r')\,
  v_{\Lambda \Lambda} (\vec{n},\vec{n}';\, r,r') ,
\label{4.15} 
\end{eqnarray}
with $v_{\Lambda \Lambda} (\vec{n},\vec{n}';\, r,r')$
where $\vec{n}=(0,0,0)$, $\vec{n}'=(a/2)(0,1,1)$ and 
$\Lambda=(l=2,\,T_{2g},k=1)$.
The coefficients $c_{ \{n_p\} }$ are defined as 
\begin{eqnarray}
  c_{ \{n_p\} }(i_{\alpha} j_{\alpha}) = 
  \langle i_{\alpha} | {\cal S}_{ \{ n_p\} } | j_{\alpha} \rangle .
\label{4.16} 
\end{eqnarray} 
We introduce the quadrupolar density operators for
the ($fd$) electron system on each sublattice:
\begin{eqnarray}
  \rho^Q_{\alpha \alpha}(\vec{n}_p) = \sum_{I,J}
   | I_{fd} \rangle  \, 
   c_{ \{n_p\} }(i_{\alpha} j_{\alpha})\, \delta_{i_{\beta} j_{\beta}} 
   \langle J_{fd} | ,
\label{4.17} 
\end{eqnarray}
where $\vec{n}_p \in \{ n_p \}$, $p=1-4$.
Here again $\alpha=\beta$, $\beta=d$ or $\alpha=d$, $\beta=f$.
In terms of quadrupolar density operators,
the quadrupolar interaction operator between two $(fd)$ systems
at site $\vec{n}_1 \in \{ n_1 \}$ and $\vec{n}_{p'} \in \{ n_{p'} \}$
reads
\begin{eqnarray}
 & &   V(\vec{n}_1, \vec{n}_{p'}) =       
  -\sum_{\alpha \alpha'} 
   \frac{\gamma^{\alpha \alpha \, \alpha' \alpha'} }{3} \, 
   \rho_{\alpha \alpha}^Q(\vec{n}_1) \, \rho_{\alpha' \alpha'}^Q(\vec{n}_{p'}) .
\label{4.18}
\end{eqnarray}
The mean-field potential at site $\vec{n}_1$ is obtained by summing 
$V(\vec{n}_1, \vec{n}_{p'})$ over the twelve nearest neighbors $\vec{n}_{p'}$
of $\vec{n}_1$ on the fcc lattice and by approximating the
quadrupolar densities at these nearest neighbor sites by their thermal
expectation values.
The thermal expectation value of $\rho_{\alpha \alpha}^Q(\vec{n}_p)$
does not depend ($i.e.$ is the same) on any site of a given sublattice and
from the equivalence of the four sublattices it follows that it is the same
on all sites of the fcc lattice. We then write
\begin{eqnarray}
 \langle \rho_{\alpha \alpha}^Q(\vec{n}_p) \rangle 
 = \rho^{Q,\,e}_{\alpha \alpha},    \label{4.19} 
\end{eqnarray}
where the superscript $e$ stand for thermal expectation.
The mean-field potential is
then given by 
\begin{mathletters}
\begin{eqnarray}
   U^{MF}(\vec{n}_1) &=& -4 \sum_{\alpha \alpha'}
   \gamma^{\alpha \alpha \, \alpha' \alpha'}
   \rho_{\alpha \alpha}^Q(\vec{n}_1) \, \rho_{\alpha' \alpha'}^{Q,\,e} ,
\label{4.20a} 
\end{eqnarray}
or, explicitly,
\begin{eqnarray}
   U^{MF}(\vec{n}_1) &=&
   -(\lambda^{F F}\,\rho^{Q,\,e}_F+\lambda^{D F}\,\rho^{Q,\,e}_D)\, 
   \rho^Q_F(\vec{n}_1)    \nonumber  \\
   & & -(\lambda^{D D}\,\rho^{Q,\,e}_D+\lambda^{F D}\,\rho^{Q,\,e}_F)\, 
   \rho^Q_D(\vec{n}_1) ,
\label{4.20b} 
\end{eqnarray}
\end{mathletters}
where we have defined $\lambda^{\alpha \alpha\, \alpha' \alpha'}
=4 \gamma^{\alpha \alpha\, \alpha' \alpha'}$ and used $F$ and $D$
for $(\alpha \alpha)$, with $\alpha=f$ or $d$, respectively.
The values of $\lambda^{\alpha \alpha\, \alpha' \alpha'}$ have
been calculated before. \cite{NM2}
Here we use the values $\lambda^{F F}=$2241 K, 
$\lambda^{D F}=\lambda^{F D}=$6489 K and $\lambda^{D D}=$18793 K,
calculated for $\alpha$-Ce.
Including the intra- site part $U|_{intra}$ and the crystal field
$V_{CF}(\vec{n}_1)$, Eq.~(\ref{4.1}), which are also present in the
ordered phase, we obtain the full mean-field Hamiltonian
\begin{eqnarray}
   H^{MF}(\vec{n}_1) = U^{MF}(\vec{n}_1) + V_{CF}(\vec{n}_1) + 
   \left. U \right|_{intra} .  
\label{4.21} 
\end{eqnarray}
Finally, the expectation values (the order parameter amplitudes)
$\rho^{Q,\,e}_F$ and $\rho^{Q,\,e}_D$, Eq.\ (\ref{4.19}), 
are found by solving the following mean-field equations
\begin{mathletters}
\begin{eqnarray}
  \rho^{Q,\,e}_F={\frac{Tr\{\rho^Q_F(\vec{n}_1)\,\exp[-H^{MF}(\vec{n}_1)/T]\}}
  {Tr\{\exp[-H^{MF}(\vec{n}_1)/T]\}}} ,
\label{4.22a}  \\
  \rho^{Q,\,e}_D={\frac{Tr\{\rho^Q_D(\vec{n}_1)\,\exp[-H^{MF}(\vec{n}_1)/T]\}}
  {Tr\{\exp[-H^{MF}(\vec{n}_1)/T]\}}} .
\label{4.22b}  
\end{eqnarray}
\end{mathletters}
It is convenient to rewrite these equations in the basis 
$|K_{fd} \rangle=| k_f k_d \rangle$ where
$H^{MF}$ is diagonal,
\begin{mathletters}
\begin{eqnarray}
  \rho^{Q,\,e}_F= \frac{1}{Z}
    \sum_{K_{fd}} c_{ \{n_1 \} }^F(k_{f} k_{f})\, e^{-\epsilon_{K_{fd}}/T} ,
\label{4.23a}  \\
  \rho^{Q,\,e}_D= \frac{1}{Z}
    \sum_{K_{fd}} c_{ \{n_1 \} }^D(k_{d} k_{d})\, e^{-\epsilon_{K_{fd}}/T} ,
\label{4.23b}  
\end{eqnarray}
with
\begin{eqnarray}
   Z=\sum_{K_{fd}} e^{-\epsilon_{K_{fd}}/T} .
\label{4.23c} 
\end{eqnarray}
\end{mathletters}
Equations (\ref{4.20b})-(\ref{4.23c})
are solved self-consistently.
First, we introduce nonzero expectation values $\rho^{Q,\,e}_F$
and $\rho^{Q,\,e}_D$ in the mean-field Hamiltonian $H^{MF}$.
After this we diagonalize $H^{MF}$ and calculate new
values of $\rho^{Q,\,e}_F$ and $\rho^{Q,\,e}_D$ at a given temperature
$T$ according to Eqs. (\ref{4.22a},b). Then we use these values 
to improve the mean-field Hamiltonian (\ref{4.21}) and $etc.$ until
the input and output values of $\rho^{Q,\,e}_F$
and $\rho^{Q,\,e}_D$ converge.
The results of the numerical calculation are shown in Fig.~1.
%
\begin{figure} 
\resizebox{0.46\textwidth}{!}
{ 
 \includegraphics{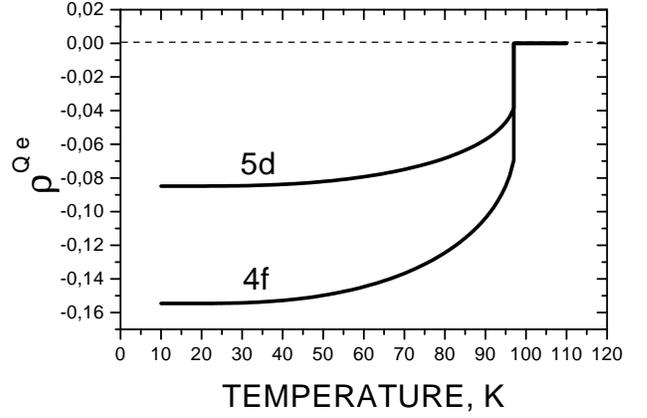} 
} 
\vspace{3mm}
\caption{
Calculated evolution of the order parameter amplitudes
$\rho^{Q,\,e}_F$ and $\rho^{Q,\,e}_D$ with temperature.
} 
\label{fig1} 
\end{figure} 
The procedure outlined above converges very slowly in the
vicinity of 100 K, $i.e.$ at the phase transition point.
We have found the transition temperature $T_1=$97 K and the
order parameter discontinuities $\rho^{Q\,e}_F(T_1)=-0.06956$
and $\rho^{Q\,e}_D(T_1)=-0.03875$.
(From symmetry
considerations it follows that the phase transition is
of first order. \cite{NM1})
At $T = 0$ the averages in Eq.~(\ref{4.19}) are taken
over the ground state doublet and we obtain
$\rho^{Q\,e}_F(T=0)=-0.15462$, $\rho^{Q\,e}_D(T=0)=-0.08485$.
The lowest five levels of $H^{MF}$ for this case are given
in Table IV.
%
\begin{table} 
\caption{ Lowest levels of the
energy spectrum of $H^{MF}$ at $T=0$ and magnetic moments ${\cal M}_z$
for $\alpha$-Ce. 
Numbers in parentheses stand for degeneracy;  
the site group is $S_6=C_3 \times i$.
\label{tab4}     } 
 
 \begin{tabular}{l c r c c } 
  $\Gamma,\;\mu$ &
 & $\epsilon_i$, in K & $(\epsilon_i-\epsilon_1)$, in K & ${\cal M}_z$, in $\mu_B$\\
\tableline
 $E$, 1 &(2) & -10888.5 &   0.0 &  $\pm$2.0683 \\
 $A$, 1 &(1) & -10721.4 &  167.1 &  0 \\
 $A$, 2 &(1) & -10699.7 &  188.8 &  0 \\
 $E$, 2 &(2) & -10542.2 &  346.3 & $\pm$1.0116 \\
 $E$, 3 &(2) & -10427.1 &  461.4 & $\pm$0.4882
 \end{tabular} 
\end{table} 
Notice, however, that unlike the crystal field $V_{CF}$ the mean field
potential $U^{MF}$, Eq.\ (\ref{4.20b}), and the Hamiltonian $H^{MF}$ 
depend implicitly on temperature $T$ since the order parameter
amplitudes $\rho^{Q\,e}_F$ and $\rho^{Q\,e}_D$ change with
temperature, Fig. 1, and the energy splittings
$(\epsilon_i-\epsilon_1)$ decrease with increasing $T$ up to $T_1$.
Above the phase transition point 
the spectrum transforms discontinuously
to that of $\gamma$-Ce, Table~III. 
The double degenerate states (the representations $E$ in Table IV)
are due to time-reversal symmetry. \cite{Tin}
For a further discussion of the energy spectrum of $H^{MF}$
we refer to Appendix~A, where we study the magnetic
moments of the two electrons.

%
%
\section {Discussion and Conclusions} 
\label{sec:con} 

This work is an extension of our previous model
of the $\gamma-\alpha$ phase transition in Ce
based on the idea of quadrupole ordering. \cite{NM1,NM2}
In addition to the inter- site quadrupolar couplings
here we consider the multipolar intra- site 
(direct Coulomb and exchange) interactions between
one localized 4$f$ electron and one delocalized 5$d$
electron taken instantaneously at a same cerium site.
The intra- site interactions are treated exactly
in the adopted $4f5d$ model.
In $\gamma$-Ce we have calculated and 
analyzed the crystal electric field excitations, Table III. 
In $\alpha$-Ce the $Pa{\bar 3}$ quadrupolar ordering sets in
and drives the $\gamma-\alpha$ phase transition.
The quadrupolar order has been studied in the mean-field approximation, 
Eqs from (\ref{4.20b}) to (\ref{4.23c}).
We have calculated the phase transition temperature ($T_1=97$ K)
and the evolution of the order parameter amplitudes 
($\rho^{Q\,e}_F$ and $\rho^{Q\,e}_D$, see Fig.~1) 
by solving self-consistently the mean-field equations 
(\ref{4.20b})-(\ref{4.23c}).
We have shown before \cite{NM1,NM2} that quadrupolar ordering
in the $Pa{\bar 3}$ structure leads to a uniform lattice contraction
conserving the cubic symmetry of the lattice.

The change of the energy spectrum at the transition
implies a change of the magnetic susceptibility.
Indeed, we have found that the calculated magnetic moments
are different in $\gamma-$ and $\alpha-$Ce. Moreover,
in the ground state of $\alpha-$Ce the magnetic
moment of the 4$f$ electron is bound to the magnetic moment of 
the 5$d$ electron 
(a qualitative origin of this correlation is given in Appendix~B).
The lowest magnetic excited state ($E$, 2 in Table IV) is separated
from the ground state by an energy gap $\triangle \epsilon \sim 350$ K
which is much larger than a typical crystal field excitation in
the $\gamma$ phase, Table III.
However, here our treatment is incomplete.
Although the present model carefully takes into account
the intra- site interactions it does not describe
properly the metallic bonding in Ce. 

The question of correspondence between our approach and electron
band structure calculations deserves a special attention.
As we have discussed in Sec.~II, in the ``muffin-tin" approximation
a localized 4$f$ electron experiences only a field of spherical
symmetry and occupies a 14-fold degenerate level.
The localized states of the 4$f$ electron then are uncorrelated
with the states of conduction electrons, because the spherical
component of the 4$f$ density is independent of its spin ($s_z$) and 
orbital ($m_f$) projections.
In our study we show that this simple picture is not correct and 
there exist strong local correlations between localized 4$f$ and delocalized
5$d$ electrons omitted in a conventional band structure calculation. 
These correlations arise due to the Coulomb 
on-site repulsion and reflect the electronic term structure of 
atomic cerium, Sec.~III. We show that the
excitations of the 4$f$ electron are combined with those of 5$d$ electron 
in a single spectrum, which is sensitive to
crystal site symmetry because of intersite interactions, Sec.~IV.

In principle, band structure calculations with the full potential (FP) 
extension (so-called FP-linear muffin-tin (FLMTO) \cite{LMTO1} 
and FP-linear augmented plane wave (FLAPW) \cite{LAPW} methods) 
are capable of dealing with non-spherical contributions of density and potential.
Provided that the site symmetry is introduced explicitly,
calculations with the full potential option can describe some, but not all
structural properties associated with the $Fm{\bar 3}m \rightarrow Pa{\bar 3}$
transformation. The reason is that 
the band structure calculations are based on the single-determinant 
Hartree-Fock method.
In our treatment each local two-electron basis function, 
Eqs.\ (\ref{2.2})-(2.6), corresponds to a Slater determinant 
(with the permutation property (\ref{2.7})). 
The solutions are expressed as linear combinations of all these functions 
(determinants). 
As such, our method corresponds to a many determinant treatment or 
configurational interaction (CI). Clearly, it is not the case with the 
band structure approach. This explains why
the conventional band structure treatment is missing some 
intra- and inter- site correlations which are 
taken into account in our approach. 

The other drawback of the FP electronic
band structure calculations is connected with the local density
approximation (LDA). In the LDA the exchange potential 
is a function of density and has the full (unit) symmetry of 
the crystal. This implies that
the density and the LDA exchange potential are invariant under inversion in
both phases. 
Expanding the exchange potential in terms of spherical harmonics
on a cerium site we find that only harmonics with even $l$ are 
allowed by the inversion symmetry, i.e., $l=2,4$ etc.
However, as we observe from Eq.\ (\ref{3.10b}) the contributions 
with {\it odd} values of $l$ are relevant for the exchange between
4$f$ and 5$d$ electrons, namely $l=1$, 3 and 5.
Thus, these terms are washed away by the LDA treatment. 

On the other hand, the present method should be extended to
include the metallic bonding in Ce explicitly. 
In our opinion, a combination of the local correlations with
the band structure approach constitutes a challenge for further studies.
We are presently
working on the problem and hope to achieve this by
employing the valence bond (VB) or Heitler-London approach.
The VB method is more difficult to implement, but usually it gives 
a better description of chemical bonding than the method of molecular
orbitals. \cite{Tin,Wil} The alternative approach is a merger with one of the
existing band structure methods and introducing in some way 
a CI treatment.

Finally, we would like to mention again that we are aware of the fact that 
our approach is not complete, but it certainly underlines the importance
of the structural factors and the local correlations for this long-standing 
problem. 
We have shown that the local electronic interactions can 
trigger the $\gamma \rightarrow \alpha$
phase transition in Ce, but other aspects of the problem (in particular, 
chemical bonding etc.) should also be taken into account. 
As before, \cite{NM1,NM2}
we suggest synchrotron radiation experiments in order to check
the appearance of weak superstructure reflections in the $Pa{\bar 3}$ 
structure ($\alpha$-Ce) and to study diffuse
scattering in $\gamma-$ and $\alpha-$Ce.

The present model
generalizes our initial approach, Ref.\ \onlinecite{NM1},
for a many electron case. 
The method can be applied 
(as it has been done for TmTe in Ref.\ \onlinecite{rem3}) 
to study quadrupole orderings \cite{Mor1} in lanthanides and their compounds
(DyB$_2$C$_2$, \cite{Dy1} DyB$_6$, \cite{Dy2} 
UCu$_2$Sn, \cite{UCu2Sn} PrPb$_3$, \cite{PrPb3} YbAs, \cite{YbAs} 
YbSb \cite{YbSb})
where a few $f$ electrons at each site are involved in the process of
ordering.

\acknowledgments 
We thank D.V. Lopaev for references on atomic spectra.
This work has been financially supported by
the Fonds voor Wetenschappelijk Onderzoek, Vlaanderen.


\appendix 

\section{}
\label{sec:apA}

In order to calculate the effective magnetic moments and the Lande $g$-factors
we study the polarization of electronic states in a small
external magnetic field $H$. 
In such case we add to the Hamiltonian ${\cal H}$
(that is $U|_{intra}$, Eq.\ (\ref{3.16}), for the atomic case,
$U|_{intra}+V_{CEF}$, Eq.~(\ref{4.11}), for $\gamma$-Ce,
and $U|_{intra}+V_{CEF}+U^{MF}$,
Eq.~(\ref{4.21}), for $\alpha$-Ce)
a magnetic term 
\begin{mathletters}
\begin{eqnarray}
   V_{mag} = - {\cal M}_z \cdot H .
\label{5.1a} 
\end{eqnarray}
Here
\begin{eqnarray}
  {\cal M}_z = ({\cal M}_z(f) + {\cal M}_z(d))  ,
\label{5.1b} 
\end{eqnarray}
\end{mathletters}
with
\begin{mathletters}
\begin{eqnarray}
  & & \vec{{\cal M}}(f) = \mu_B (\vec{L}(f) + 2 \vec{S}(f)) , \label{5.2a} \\
  & & \vec{{\cal M}}(d) = \mu_B (\vec{L}(d) + 2 \vec{S}(d)) , \label{5.2b} 
\end{eqnarray}
\end{mathletters}
$\mu_B$ being the Bohr magneton. The matrix elements of $V_{mag}$ read
\begin{eqnarray}
 \langle I_{fd} | V_{mag} | J_{fd} \rangle &=&
 (\langle i_f | {\cal M}_z(f) | j_f \rangle\, \delta_{i_d j_d}  \nonumber  \\
 &+& \langle i_d | {\cal M}_z(d) | j_d \rangle\, \delta_{i_f j_f}) \cdot H ,  
\label{5.3} 
\end{eqnarray}
where $\langle i_f | {\cal M}_z(f) | j_f \rangle$ 
and $\langle i_d | {\cal M}_z(d) | j_d \rangle$ are one-particle
matrix elements which can be easily computed.
If we diagonalize the matrix of $({\cal H}+V_{mag})$, then
the degeneracies of  the energy terms
are lifted and the magnetic moment of
each sublevel $\nu$ is given by 
\begin{eqnarray}
  {\cal M}_z(\epsilon_{\nu}) = 
  \langle \epsilon_{\nu} | {\cal M}_z | \epsilon_{\nu} \rangle ,
\label{5.4} 
\end{eqnarray}
where $\{ \epsilon_{\nu} \}$ stands for the energy levels $\{ E_{fd} \}$
or $\{ \epsilon_i \}$ given in Tables II or III and IV, for the atomic case
(intra- site), the $\gamma$-phase and the $\alpha$-phase, respectively.
The results for ${\cal M}_z$ for the solid state phases are quoted in
Tables III and IV.
The Lande $g$-factors are obtained from
\begin{eqnarray}
  {\cal M}_z(\epsilon_{\nu},j) = \mu_B\, g(\epsilon_{\nu})\, j  ,
\label{5.5} 
\end{eqnarray}
where $j$ is the $z$-projection of a total angular moment:
$j=-J,-J+1,...,+J$. The calculated $g$-factors for the atomic case, Table II,
are close to the experimental values. \cite{at_sp}

In case of $\gamma$-Ce and $\alpha$-Ce the $z$-axis is the [001] axis
of the cubic crystal.  For $\gamma$-Ce only triply degenerate
levels of $T_1$ or $T_2$ symmetry have nonzero magnetic moments which
are $-{\cal M}(t)$, 0 and $+{\cal M}(t)$, where ${\cal M}(t)>0$ and $t=(T_1,\mu)$
or $t=(T_2,\mu)$, Table III.
For $\alpha$-Ce only double degenerate levels of $E$ symmetry
have nonzero magnetic moments which are $-{\cal M}(e)$ and $+{\cal M}(e)$,
where ${\cal M}(e)>0$ and $e=(E,\mu)$, Table IV.
The double degenerate states 
are due to time-reversal symmetry. \cite{Tin} 
It is worth noting that in the ground state ($E$,1) the magnetic moment
of the localized 4$f$ electron is not independent.
It is attached to the magnetic moment
of the 5$d$ conduction electron. The
change of sign of the magnetic moment of the 5$d$ electron 
under the time reversal symmetry now requires
the concomitant change of the sign of the magnetic moment of the
localized 4$f$ electron. 
We discuss the
origin of this correlation in Appendix B. 
Since the 5$d$ electron belongs to the conduction band,
the ground state ($E$,1) gives rise to a Pauli paramagnetism.
The other levels given in Table~IV represent
excited states of $\alpha$-Ce. However, we observe that
the first (representation $A$,1 of Table IV) and the
second state ($A$,2) are nonmagnetic. The first magnetic
excited state ($E$,2) is separated from the ground state
by an energy gap of $\triangle \epsilon \sim 350$ K. 
This observation could explain the absence of the Curie-Weiss
contribution to the magnetic susceptibility $\chi_{\alpha}$
at temperatures $T<T_1$.

\section{} 
\label{sec:apB}

We consider the function ${\cal S}_{ \{n_1\} }(\Omega)$, Eq.\ (\ref{4.12'a}),
in a new coordinate system of axes $\{ x',y',z' \}$ where the $z'$-axis
corresponds to the cubic [111]-axis, Fig.~2.
%
\begin{figure} 
\resizebox{0.40\textwidth}{!}
{ 
\hspace{1cm} \includegraphics{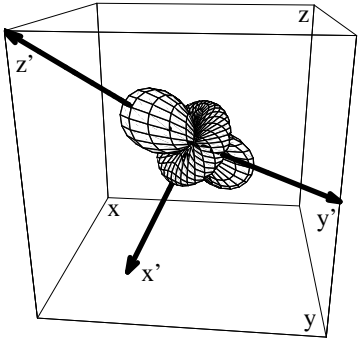} 
} 
\vspace{3mm}
\caption{
The transformation $\{x,y,z \} \rightarrow \{x',y',z' \}$
(the Euler angles are $\alpha=\pi/4$, $\beta=arccos(1/\sqrt{3})$, $\gamma=0$
for passive and $(\gamma,\beta,\alpha)$ for active rotations), 
and the function ${\cal S}_{ \{n_1\} }$, 
Eq.~(\ref{4.12'a}).
} 
\label{fig2} 
\end{figure} 
Then
\begin{eqnarray}
   {\cal S}_{ \{n_1\} }(\Omega)=Y_{l=2}^{m=0}(\Omega')  ,
\label{b.1} 
\end{eqnarray}
where $\Omega'$ stands for the polar angles $(\Theta',\phi')$ in the new
system of axes.
The coefficients $c^F_{ \{ n_1\} }(i_f j_f)$ and
$c^D_{ \{ n_1\} }(i_d j_d)$, Eq.\ (\ref{4.16}), 
become diagonal in the new basis,
\begin{eqnarray}
  c_{ \{ n_1\} }(i_{\alpha} j_{\alpha})=
  \langle i_{\alpha} | Y_2^0 | i_{\alpha} \rangle \,
  \delta_{i_{\alpha} j_{\alpha}}   ,
\label{b.2} 
\end{eqnarray}
that facilitates the calculation of the mean-field interaction
$U^{MF}(\vec{n}_1)$, Eq.~(\ref{4.20b}).
In this Appendix we show that $U^{MF}$
is minimized for the following two functions
\begin{mathletters}
\begin{eqnarray}
 & & {\cal Y}_{L=5}^{M_L=5}=Y_3^3(\Omega'_f) \cdot Y_2^2(\Omega'_d) , 
 \label{b.3a} \\ 
 & & {\cal Y}_{L=5}^{M_L=-5}=Y_3^{-3}(\Omega'_f) \cdot Y_2^{-2}(\Omega'_d) , 
  \label{b.3b} 
\end{eqnarray}
\end{mathletters}
where $\Omega'_f$ and $\Omega'_d$ refer to the polar coordinates
of 4$f$ and 5$d$ electron, respectively.

Indeed, there are 11 functions ${\cal Y}_{L=5}^{M_L}$
($M_L=-5,-4,...,5$), which form a basis of the 
two-particle irreducible representation $H$ of the group $SO(3)$ of three
dimensional rotations. 
(In Eq.\ (B3) we quote only two functions, the others can be
obtained from the Table 4$^3$ of Ref.\ \onlinecite{CS}.) 
In the $LS$ (Russel-Saunders)
coupling the triplet $^3 H$ lies lower in energy
than the singlet $^1 H$, Table I. If now we include the 
diagonal mean-field
coupling $U^{MF}$, we obtain that the six states
with the orbital components ${\cal Y}_{5}^{5}$ and ${\cal Y}_{5}^{-5}$
and with the spin components $M_S=-1$, 0, 1 (originating from the
$^3 H$ triplet) go down in energy. These six states are further split
by the spin-orbit coupling in three magnetic doublets of $E$
symmetry of the site group $S_6=C_3 \times i$. 
Finally, in the full treatment (Sect.~4.2) the middle doublet
is split by the crystal field in two nonmagnetic components of $A$
symmetry. This explains qualitatively the origin of the four lowest levels
of the $Pa{\bar 3}$ structure of $\alpha$-Ce, Table IV.
The real situation is more complex since there is a mixing 
of different configurations in atomic cerium. 
Nevertheless, a considerable admixture
of $^3 H$ configuration (29\%) is found in the ground state
of atomic cerium \cite{at_sp} and it is very likely to occur
in $\gamma$-Ce.

In $\alpha$-Ce orbital functions of 5$d$ electrons on neighboring sites 
overlap giving rise to band structure effects.
Here we consider these effects only as a perturbation to the ground state
$E,1$, Table IV. Then in tight-binding approximation 
a band state with the wave vector $\vec{k}$
is a mixture of the $d-$functions $Y_2^2(\Omega'_d)$ and $Y_2^{-2}(\Omega'_d)$.
However, as we have seen earlier, 
these electronic 5$d$ states are bound to 
the two 4$f$ electron states $Y_3^3(\Omega'_f)$
and $Y_3^{-3}(\Omega'_f)$ and form the two-electron functions
${\cal Y}_{L=5}^{M_L=5}$ and ${\cal Y}_{L=5}^{M_L=-5}$, Eqs.\ (\ref{b.3a},b).
Therefore, the resulting doubly degenerate states are in fact 
two-electron band states with the energy 
${\cal E}_+(\vec{k})={\cal E}_-(\vec{k})$ 
(the sign $+$ or $-$ here stands for the two-electron 
time reversed band states).
In the magnetic field these two bands become split,
 ${\cal E}_+(\vec{k}) \neq {\cal E}_-(\vec{k})$, 
and this leads
to a temperature independent Pauli paramagnetism.



\begin{references}
\bibitem{rem2}
The values of $B_4^d$, $B_4^f$ and $B_6^f$ were
calculated taking into account all neighbors, see Ref.\ \onlinecite{NM3}.
\end{references}
\end{document}